*Raman Topography and Strain Uniformity of Large-Area Epitaxial Graphene*

J. A. Robinson,[1,2,*] C. P. Puls,[3] N. E. Staley,[3] J. Stitt,[2] M.A. Fanton,[1] K. V. Emtsev,[4] T. Seyller[4], & Y. Liu[2,3]

[1] Electro-Optics Center, The Pennsylvania State University, 559A Freeport Rd., Freeport, PA 16229
[2] Materials Research Institute, The Pennsylvania State University, 230 Innovation Blvd., University Park, PA 16802
[3] Department of Physics, The Pennsylvania State University, University Park, PA 16802
[4] Lehrstuhlfür Technische Physik, Universität Erlangen-Nürnberg, Erwin-Rommel-Strasse 1, 91058 Erlangen, Germany

**We report results from two-dimensional Raman spectroscopy studies of large-area epitaxial graphene grown on SiC. Our work reveals unexpectedly large variation in Raman peak position across the sample resulting from inhomogeneity in the strain of the graphene film, which we show to be correlated with physical topography by coupling Raman spectroscopy with atomic force microscopy. We report that essentially strain free graphene is possible even for epitaxial graphene.**

Graphene exhibits extraordinary electronic properties including an unusually high mobility of the charge carriers.[1] While significant progress toward understanding the properties of graphene has resulted from studying graphene flakes mechanically exfoliated from bulk graphite,[2] these small flakes (< 100 $\mu m^2$) are most suited for studying the fundamental science of graphene, and are not practical for the development of graphene-based technologies.  Alternatively, the sublimation of silicon (Si) from silicon carbide (SiC) to form epitaxial graphene is a promising route for the production of wafer size graphene films.[3-9] However, rapid characterization and precise control of properties of epitaxial graphene over a wafer-size area are yet to be achieved. Micro-Raman spectroscopy is a rapid, high-resolution optical characterization technique that yields important information on the thickness, the charge carrier density, and the strain of epitaxial graphene.[10,11,12,13] However, no studies of Raman topography, the two-dimensional mapping of Raman spectrum over large-area epitaxial graphene, have been carried out to date.

---

[*] Corresponding Author: jrobinson@psu.edu

Bulk graphite and graphene exhibit defect-induced Raman signals at approximately 1360 cm$^{-1}$ (D peak), and an overtone peak near 2700 cm$^{-1}$ (2D peak).[14,15] Both signals are due to resonant Raman scattering where the specific frequency value is dependent on the laser excitation energy,[16] which is attributed to a double resonant Raman process near the K-point of the Brillouin zone.[15] In this Letter we report that monolayer graphene films appear to exhibit large variation in the 2D peak position, and include unexpected values below that of bulk graphite. We show that this is due to strain relief via mechanical decoupling from the substrate, thus reducing the film strain to values similar to exfoliated graphene.

Monolayer epitaxial graphene was synthesized via Si-sublimation from the Si-face of semi-insulating 6H-SiC (0001) as described in detail elsewhere.[6] The thickness was independently verified by photoemission spectroscopy.[17] A WITec confocal Raman microscope (CRM) with a 488 nm laser wavelength, diffraction limited lateral resolution of ~ 340 nm, and spectral resolution of 0.24 cm$^{-1}$ was utilized for Raman spectroscopy. An example of a full Raman spectrum obtained on our epitaxial graphene films is presented in the supplementary material. We choose to focus on the 2D peak in the Raman spectrum in graphene due to concerns regarding subtraction of the SiC background seen in, for example, the G peak.[12] For Raman topography we map a two-dimensional region of a graphene film, collecting Raman spectra with a step size of 300 nm in both x and y directions. The position of each Raman peak was identified using the center of mass of the peak via the equation $\sum_i I_i \omega_i / \sum_i I_i$, where and $\omega_i$ and $I_i$ are the spectral wavenumber and the intensity Raman signal at position i in the x-y plane. To simplify the computation, we focus on a range of 2650 – 2825 cm$^{-1}$ at each sample location (x,y) including only the 2D peak. Layer thickness was identified via a Lorentzian fit. Spectra that can be fitted to a single Lorentzian are mono- and those fitted to the sum of four Lorentzians are bilayer graphene.[18] The physical topography of the graphene film was determined by atomic force microscopy using a Digital Instruments Nanoscope 3A and correlated with the two-dimensional Raman map of the same location.

The Raman 2D peak position is used to infer the thickness of an exfoliated graphene flake, which converges to that of bulk graphite as the layer thickness increases.[12, 18, 19] In Figure 1a, the positions of the 2D peak in a line scan are plotted. Interestingly, its values are found to vary significantly, ranging from 2689 cm$^{-1}$ to 2754 cm$^{-1}$, with many values falling below that of bulk graphite.[10,12] Similarly, large deviation of the 2D peak position from that found in exfoliated graphene flakes was found previously in epitaxial graphene films.[10,11,12] The variation in the 2D peak position could be a result of graphene thickness non-uniformities. However, fitting individual spectra taken in this line scan using a single Lorentzian indicates that the film was indeed monolayer graphene as demonstrated independently by photoemission spectroscopy. Representative spectra are shown in Figure 1b taken at positions marked by dots in Figure 1a. Therefore the observed variation in the 2D peak position must be related to variations in the properties of monolayer graphene.

Variations in the 2D peak position result from charge or strain inhomogeneities of the monolayer graphene. A charge transfer from the substrate to the graphene was found previously[20,21] to be roughly 2 x 10$^{13}$/cm$^2$. A variation of the charging level due to unintentional doping and other causes on the same order of magnitude of the average carrier density itself would shift the 2D peak by small amount (≈7 cm$^{-1}$),[22] compared to the ≈64 cm$^{-1}$ shift seen in Fig. 1b. On the other hand, an increase in the 2D peak position to above that of bulk graphite was observed previously[23] and attributed to a compressive strain induced by the substantial lattice mismatch between graphene and SiC (≈20%). A large difference in the coefficient of thermal expansion will lead to further compression of the graphene film[10,12,24] upon cooling the sample from growth temperature (>1200°C) to room temperature. Because the large peak shifts seen experimentally (Fig. 1b) are greater than the 7 cm$^{-1}$ expected from the change in charge carriers, we can attribute the 2D peak shift seen in our monolayer graphene to strain effects.

As a result we can assign Raman peak values higher than bulk graphite (~ 2720 cm$^{-1}$) to high-strain,[10,12] and those below that to low-strain monolayer epitaxial graphene.

Spectra featuring a peak that can be fitted by two Lorentzians were also observed. For example, fitting selected spectra acquired in three consecutive spots revealed the presence of strain state boundaries (Fig.2). We believe that such spectra may have resulted from the finite spot size of the Raman laser (≈340nm), which would suggest that the strain state may be varying on a surprisingly short length scale (< 340 nm). Careful analysis of the Raman topographic maps indicates that in epitaxial graphene the strain may only be uniform on the micron scale. Incidentally, results shown in Fig. 2 also indicate that the graphene thickness should be determined by fitting the peak to a Lorenzian as opposed to making use of peak width.

To fully explore the physical origin of strain variation in epitaxial graphene, we introduce a combination of Raman topography and atomic force microscopy (AFM) measurements. To correlate strain variation with film morphology, we have mapped multiple regions on the SiC surface, including defect sites known as micropipe dislocations.[25] Distinct similarities were found in the Raman and AFM maps as shown in Figure 3a and b, which demonstrates that physical topography plays a key role in determining the strain uniformity of epitaxial graphene. The hexagonal topographic pattern typical in micropipes (Fig. 3a) can clearly be seen in the Raman map (Figure 3b). Similar results are also found in regions free of macro-defects (Figure 3c and d). The terrace step-edges appear to significantly influence the strain state in epitaxial graphene films.

In summary, we have shown that the substrate/graphene interaction, and thus the strain, significantly alters the position of the 2D Raman peak of monolayer graphene. As a result, peak position and peak width are not sufficient to determine the graphene thickness. However, the use of a single Lorentzian fit to the 2D Raman peak can identify monolayer graphene unambiguously. The Raman topography characterization of a large-area epitaxial graphene has been shown to yield important information not only on the graphene thickness, but also strain uniformity.  Finally, the observation of extremely abrupt change in the high and low strain regions in an epitaxial graphene was surprising. This strain variation will undoubtedly affect the electronic transport and other properties of the graphene film. Understanding and ultimately controlling the graphene/substrate interface will therefore be crucial for the realization of graphene-based technologies.

**References**

[1] Geim, A., Novoselov. The Rise of Graphene. Nature Materials 6, 183-191 (2007)

[2] Novoselov, K.S. *et al.* Electric field effect in atomically thin carbon films. *Science* **306**, *666-669* (2006).

[3] Berger, C. *et* al. Ultra thin epitaxial graphite: two-dimensional electron gas properties and a route toward graphene-based nanoelectronics. *J.Phys.Chem.B* **108**, 19912–19916 (2004).

[4] Starke, U. *et al.* Large unit cell superstructures on hexagonal SiC-surfaces studied by LEED, AES and STM. *Mater. Sci. Forum* **321**, 264-268 (1998).

[5] Hass, J. *et al*. Highly ordered graphene for two dimensional electronics. *Appl. Phys. Lett.* **89**, 143106 (2006).

[6] Seyller, T. *et al.* Structure and electronic properties of graphite layers grown on SiC(0001). *Surf. Sci.* **600**, 3906-3911 (2006).

[7] Forbeaux, I., Themlin, J.M. & Debever, J.M. High-temperature graphitization of the 6H-SiC (000$\bar{1}$) face. *Surf. Sci.* **442**, 9-18 (1999).

[8] Johansson, L.I., Glans, P.A., & Hellgren, N. A core level and valence band photoemission study 6H-SiC (000$\bar{1}$). *Surf. Sci.* **405**, 288-297 (1998).

[9] de Heer, W. A. *et al.* Epitaxial Graphene. *Solid State Comm.* **143** 92-100 (2007).

[10] Ni, Z.H. *et al.* Raman spectroscopy of epitaxial graphene on a SiC substrate. *Phys. Rev. B* **77**, 115416 (2008).

[11] Faugeras, C. *et al.* Few-layer graphene on SiC, pyrolitic graphite, and graphene: A Raman scattering study. *App. Phys. Lett.* **92**, 011914 (2008).

[12] Röhrl, J. *et al.* Raman spectra of Epitaxial Graphene on SiC (0001). *Appl. Phys. Lett.* **92**, 201918 (2008).

[13] Varchon, F. *et al.* Electronic Structure of Epitaxial Graphene Layers on SiC: Effect of the Substrate. *Phys. Rev. Lett.* **99**, 126805 (2007).

[14] Thomsen, C., Reich, S. & Maultzsch, J. Resonant Raman spectroscopy of nanotubes. *Phil. Trans. R. Soc. Lond. A* **362**, 2337–2359 (2004).

[15] Thomsen, C. & Reich, S., Double-resonant Raman scattering in graphite. *Phys.Rev.Lett.* **85**, 5214 (2000).

[16] Vidano, R.P., Fischbach, O.B., Willis, L.J. & Loehr, T.M. Observation of Raman Band Shifting with Excitation Wavelength for Carbons and Graphites. *Solid State Communications* **39**, 341-344 (1981).

[17] Emtsev, K.V., Speck, F., Seyller, T., Ley, L. & Riley, J.D. Interaction, growth, and ordering of epitaxial graphene on SiC{0001} surfaces: A comparative photoelectron spectroscopy study. P*hys. Rev. B* **77**, 155303 (2008).

[18] Graf, D. *et al.* Spatially Resolved Raman Spectroscopy of Single- and Few-Layer Graphene. *Nanoletters* **7**, 238-242 (2007).

[19] Ferrari, A.C. *et al.* Raman Spectrum of Graphene and Graphene Layers. *Phys. Rev. Lett.* **97**, 187401 (2006).

[20] Ohta, T. A. Bostwick, T. Seyller, K. Horn, and E. Rotenberg. Controlling the elctronic structure of bilayer graphene. Science 313 951 (2006).

[21] Ohta, T. *et al.* Interlayer Interaction and Electronic Screening in Multilayer Graphene Investigated with Angle-Resolved Photoemission Spectroscopy. *Phys. Rev. Lett.* **98**, 206802 (2007).

[22] Das, A. *et al.* Electrochemically Top Gated Graphene: Monitoring Dopants by Raman Scattering. Nature Nanotechnology **3**, 210 (2008).

[23] This value is based on $a_{SiC}$ = 3.073Å, $a_{graphene}$ = 2.456Å, and $\Delta a = (a_{SiC} - a_{graphene})/a_{SiC}$

[24] Ferralis, N., Maboudian, R., Carraro, C. Evidence of Structural Strain in Epitaxial Graphene Layers on 6H-SiC(0001). Preprint at <http://arxiv.org/abs/arXiv:0808.3605v1> (2008)

[25] Muller, St. G et al., Large area SiC substrates and epitaxial layers for high power semiconductor devices — An industrial perspective. Superlattices and Microstructures **40,** 200 (2006).

**Acknowledgements**. The WiteC Raman system used in the present work is part of the National Nanotechnology Infrastructure Network at Penn State. J.A.R. would like to acknowledge discussions with J.T. Robinson at the Naval Research Laboratory. C.P.P., N.E.S. and Y.L. were supported by DOE under Grant DE-FG02-04ER46159 and DOD ARO under Grant W911NF-07-1-0182.

**Figure Captions**

**Figure 1.** **a**) Raman 2D peak shift as a function of lateral position obtained in a Raman line scan over a monolayer epitaxial graphene. The variation of the center of mass (see text) peak position is approximately 65 cm$^{-1}$; **b)** Individual spectra obtained in positions as indicated by dots shown in (a) and the fits using a single Lorentzian. The excellent fit indicates that the graphene is indeed a monolayer graphene.

**Figure 2.** Three consecutive spectra collected in a line scan (inset). The changes in the lateral position for individual spectra are indicated. The fit to a single Lorentzian in the top and bottom curves demonstrates that the film is monolayer graphene. The middle curve is fitted to a convolution of the same Lorentzians in the two adjacent points in this Raman scan. The fit suggests that the middle Raman spectrum is collected over two areas of monolayer graphene subjected to different strain. An abrupt transition in the strain is evident.

**Figure 3.** A comparison of an AFM and Raman spectral map of the 2D peak position (**a - b**) near a SiC micropipe defect where the location of the micropipe is marked with an "x" and (**c - d**) where such a defect is not present. The Raman topography is seen to be correlated with the physical topography of the graphene film as revealed by AFM, suggesting that changes in the physical topography may lead to corresponding changes in the strain of the graphene film.

**Figure 1**

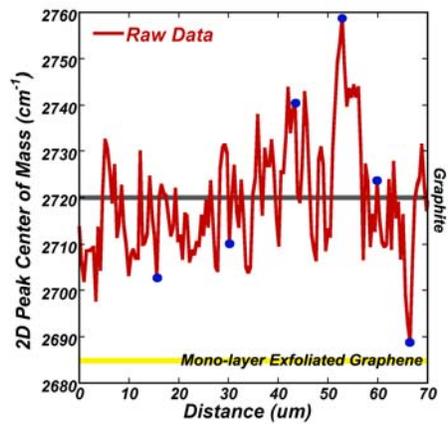

(a)

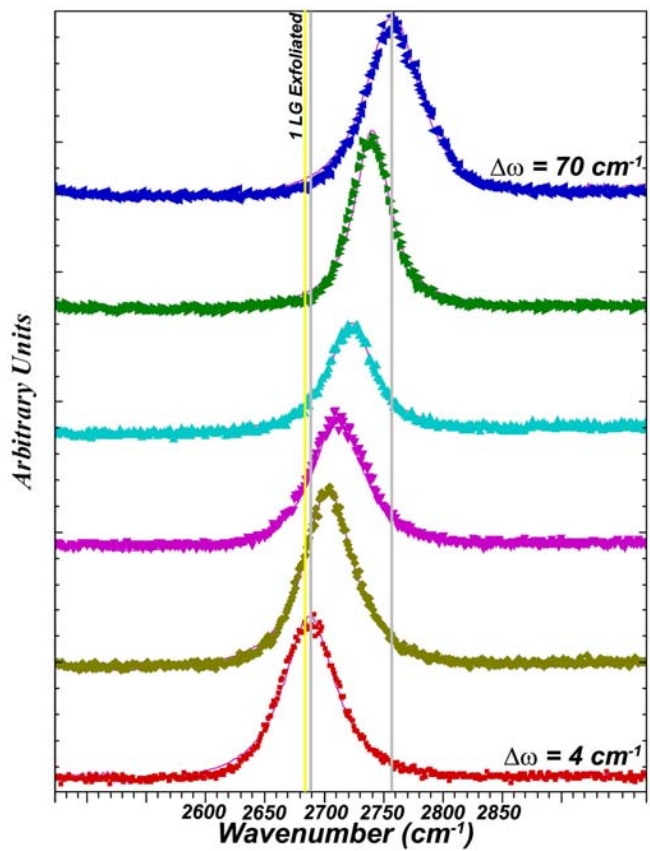

(b)

**Figure 2**

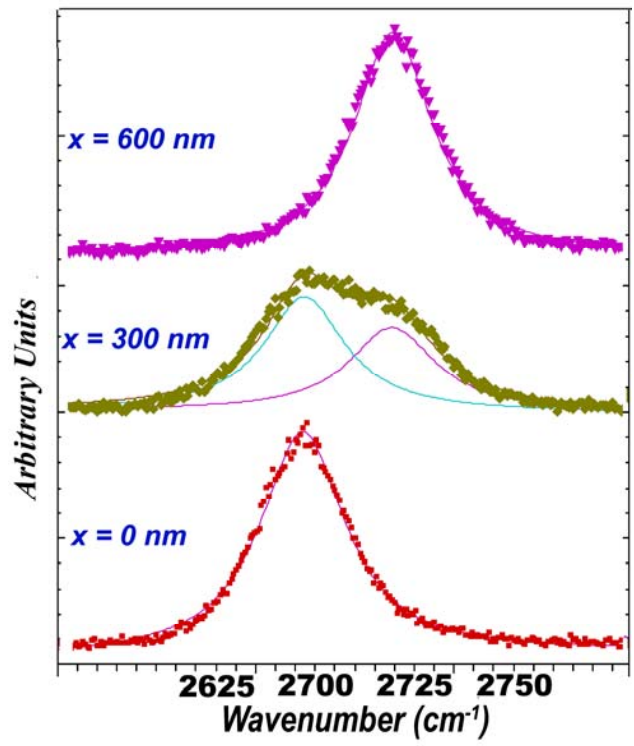

Figure 3

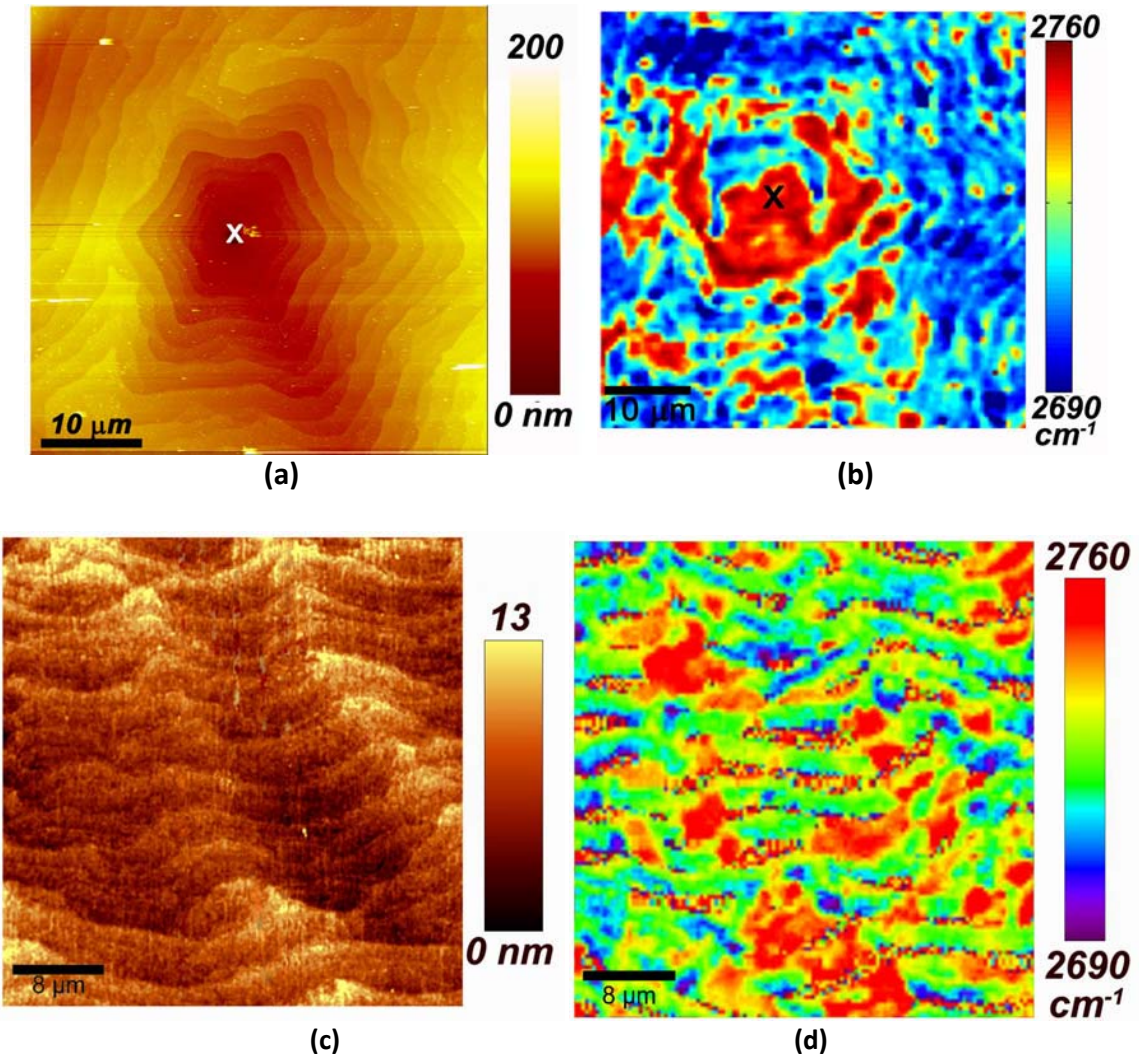